\documentclass[12pt,letterpaper]{article}

\usepackage{scicite}

\usepackage{amsmath,amssymb,amsfonts,amsthm} 

\usepackage{graphicx}
\usepackage{times}
\usepackage{color}
\usepackage{subfigure}
\usepackage{setspace} %% for singlespacing
\usepackage{url}
\usepackage{ifthen}

\newboolean{astroph}
\setboolean{astroph}{true}

\newenvironment{sciabstract}{%
\begin{quote} \bf}
{\end{quote}}

\newcounter{lastnote}

\newcommand\startsupplement{%
    \makeatletter 
       \setcounter{table}{0}
       \renewcommand{\thetable}{S\arabic{table}}
       \setcounter{figure}{0}
       \renewcommand{\thefigure}{S\arabic{figure}}
	   \renewcommand{\thesection}{S\arabic{section}}
    \makeatother}

\newcommand{\CII}{\hbox{{\rm C}\kern 0.1em{\sc ii}}}
\newcommand{\CIV}{\hbox{{\rm C}\kern 0.1em{\sc iv}}}
\newcommand{\Fe}{\hbox{{\rm Fe}}}
\newcommand{\NV}{\hbox{{\rm N}\kern 0.1em{\sc v}}}
\newcommand{\FeI}{\hbox{{\rm Fe}\kern 0.1em{\sc i}}}
\newcommand{\FeII}{\hbox{{\rm Fe}\kern 0.1em{\sc ii}}}
\newcommand{\Si}{\hbox{{\rm Si}}}
\newcommand{\Cr}{\hbox{{\rm Cr}}}
\newcommand{\SiII}{\hbox{{\rm Si}\kern 0.1em{\sc ii}}}
\newcommand{\AlII}{\hbox{{\rm Al}\kern 0.1em{\sc ii}}}
\newcommand{\NiII}{\hbox{{\rm Ni}\kern 0.1em{\sc ii}}}
\newcommand{\CrII}{\hbox{{\rm Cr}\kern 0.1em{\sc ii}}}
\newcommand{\Zn}{\hbox{{\rm Zn}}}
\newcommand{\ZnH}{\hbox{\Zn/{\rm H}}}
\newcommand{\ZnII}{\hbox{{\rm Zn}\kern 0.1em{\sc ii}}}
\newcommand{\NII}{\hbox{[{\rm N}\kern 0.1em{\sc ii}}]}
\newcommand{\OVI}{\hbox{{\rm O}\kern 0.1em{\sc vi}}}
\newcommand{\OIII}{\hbox{[{\rm O}\kern 0.1em{\sc iii}}]}
\newcommand{\OII}{\hbox{[{\rm O}\kern 0.1em{\sc ii}}]}
\newcommand{\OI}{\hbox{[{\rm O}\kern 0.1em{\sc i}}]}
\newcommand{\OH}{\hbox{{\rm O}/{\rm H}}}
\newcommand{\MgI}{\hbox{{\rm Mg}\kern 0.1em{\sc i}}}
\newcommand{\MgII}{\hbox{{\rm Mg}\kern 0.1em{\sc ii}}}

\newcommand{\HI}{\hbox{{\rm H}\kern 0.1em{\sc i}}}
\newcommand{\HII}{\hbox{{\rm H}\kern 0.1em{\sc ii}}}
\newcommand{\lya}{\hbox{{\rm Ly}\kern 0.1em$\alpha$}}
\newcommand{\Ly}{\hbox{{\rm Ly}\kern 0.1em$\alpha$}}
\newcommand{\Lyb}{\hbox{{\rm Ly}\kern 0.1em$\beta$}}
\newcommand{\Lyd}{\hbox{{\rm Ly}\kern 0.1em$\delta$}}
\newcommand{\Lye}{\hbox{{\rm Ly}\kern 0.1em$\epsilon$}}

\newcommand{\Ha}{\hbox{{\rm H}\kern 0.1em$\alpha$}}
\newcommand{\Hb}{\hbox{{\rm H}\kern 0.1em$\beta$}}

\newcommand{\flux}{erg~s$^{-1}$~cm$^{-2}$}

\newcommand{\mpy}{\hbox{$M_{\odot}$~year$^{-1}$}}

\newcommand{\msun}{\hbox{$M_{\odot}$}}
\newcommand{\cmsq}{\hbox{cm$^{-2}$}}
\newcommand{\NH}{\hbox{$N_{\rm H}$}}
\newcommand{\NHI}{{\ensuremath{N_\textsc{h\,\scriptsize{i}}}}}

\newcommand{\arcsec}{\hbox{$^{\prime\prime}$}}

\newcommand{\logZ}{\log Z/Z_{\odot}}

\newcommand{\kms}{km~s$^{-1}$}
\newcommand{\nn}{\nonumber}

\def \aap{A\&A}

\def \apjl{ApJ}
\def \apjs{ApJS}
\def \apj{ApJ}

\def \mnras{MNRAS}

\def \nat{Nat}

\def \pasp{PASP}

\newcommand{\sfr}{$33^{+40}_{-11}$~\mpy}
\newcommand{\zHa}{ 2.3283} 

\newcommand{\Zndla}{-0.72}
\newcommand{\degree}{\ensuremath{^\circ}}

\title{Signatures of Cool Gas Fueling a Star-Forming Galaxy at Redshift 2.3}
\author
{
N. Bouch\'e$^{1,2,\ast}$, M. T. Murphy$^{3}$, G. G. Kacprzak$^{3}$ , C. P\'eroux$^{4}$,\\
T. Contini$^{1,2}$, C. L. Martin$^5$, M. Dessauges-Zavadsky$^{6}$\\
\\
\normalsize{$^1${CNRS/IRAP, 14 Avenue E. Belin, F-31400 Toulouse, France}}\\
\normalsize{$^2${University Paul Sabatier of Toulouse/ UPS-OMP/ IRAP, F-31400 Toulouse, France }}\\
\normalsize{$^3${Swinburne University of Technology,   PO Box 218, Hawthorn, Victoria 3122, Australia}}\\
\normalsize{$^4${Aix Marseille Universit\'e, CNRS, LAM (Laboratoire d'Astrophysique de Marseille) UMR 7326,}}\\ \normalsize{F-13388  Marseille, France}\\
\normalsize{$^5${Department of Physics, University of California, Santa Barbara, CA 93106, USA}}\\
\normalsize{$^6${Observatory of Geneva, 51  chemin des Maillettes, CH-1290 Versoix, Switzerland}}\\
\normalsize{$^\ast$To whom correspondence should be addressed. E-mail: nicolas.bouche@irap.omp.eu.}
}

\date{}

\begin{document}


\baselineskip24pt


\maketitle


\begin{sciabstract}
Galaxies are thought to be fed by the continuous  accretion of intergalactic gas,
 but  direct observational evidence has been elusive. 
The accreted gas is expected to orbit about the galaxy's halo, 
delivering not just fuel for star-formation but also angular momentum 
to   the galaxy, leading to distinct kinematic signatures.
We report observations showing these distinct signatures  near a typical distant star-forming galaxy  
where the gas is detected using  a  background quasar passing 26~kpc   from the host.
Our observations  indicate that gas accretion plays  a major role in galaxy growth 
since the estimated accretion rate is comparable to the star-formation rate.
\end{sciabstract}

At all epochs, galaxies have short gas depletion time scales \cite{DaddiE_10a,GenzelR_10a};
to sustain the observed levels of star-formation over many billions of years, 
galaxies must continuously replenish their gas reservoir with fresh gas 
accreted from the vast amounts available in the intergalactic medium.
In numerical cosmological simulations \cite{KeresD_05a,FaucherG_11b,vandeVoortF_11a}, the accretion phenomenon is often referred to as `cold accretion' \cite{DekelA_09a}  and this term describes the mass regime where the accretion is most efficient \cite{WhiteS_91a,BirnboimY_03a}. %Such simulations  and  simple arguments \cite{GenelS_08a,DekelA_09a,BoucheN_10a,DuttonA_10a}  
%indicate that the accretion rate should be comparable to the star-formation rate (SFR).
The cold accreted gas should orbit about the halo before falling in to build the central disk, delivering fuel for star formation and also angular momentum to shape the outer parts of the galaxy  \cite{StewartK_11b,ShenS_13a}. Thus, accreting material should co-rotate with the central disk in the form of a warped, extended cold gaseous disk, producing distinct kinematic signatures in absorption systems. In particular,
the gas kinematics are expected to be offset by about $100$~\kms\ from the galaxy's systemic velocity and these  kinematic signatures of gas accretion  should be observable 
in suitable  quasar absorption line systems \cite{DekelA_09a,StewartK_11a,FumagalliM_11a,GoerdtT_12a}.  

Here,
we describe observations of a background quasar whose apparent location on the sky is fortuitously aligned with the galaxy’s projected major-axis, making it possible to test these predictions
The associated star-forming galaxy with redshift $z=2.3285$ is located just 26~kpc from the damped Lyman absorber (DLA) seen towards the quasar HE 2243$-$60 \cite{LopezS_02a}.
The galaxy was detected in our  $z\simeq2$ SINFONI~\cite{sinfo} survey called the  SINFONI \MgII\ Program for Line Emitters  (SIMPLE) \cite{BoucheN_11a}.
Recent adaptive optics (AO) assisted SINFONI observations \cite{SOM} of this $z=2.3285$ star-forming galaxy  (Fig.~1a)
obtained at the Very Large Telescope (VLT) with $\sim1$~kpc ($0.25$ arc~sec) resolution  (table S1) allow us to map the emission kinematics with precision (Fig.~1b and figs S3 and S4).
The kinematics reveal that this  galaxy has  physical properties  (Table~\ref{table:summary})
 typical for  rotationally-supported disks seen at that redshift \cite{ForsterSchreiberN_09a}.
For instance, the galaxy has a star-formation rate (SFR) of \sfr\ (where $\msun$ is the mass of the Sun), its maximum rotation velocity is $150\pm15$~\kms\ from 3D fitting (Fig.~S4, Table~S3), and its metallicity is about 1/2 solar  ($[\OH]=-0.35\pm$0.1 dex) determined from a joint fit to all the major nebular emission lines (\OII, \OIII$+$\Hb\ and \Ha/\NII; figs.~S1 and S2).
% placing the galaxy on the normal mass-metallicity scaling relations \cite{ErbD_06a}.

Analysis of a deep high-resolution VLT/UVES (Ultraviolet and Visual Echelle Spectrograph) spectrum of the background quasar HE 2243$-$60 (fig.~S5) \cite{SOM} shows that the  gas metallicity  can give us insights into the physical nature of the gas.
In particular, the total \HI\ column  is $\log(\NH/\cmsq)\simeq 20.6 $ (i.e. almost entirely neutral) and, from the undepleted low-ionisation ion \ZnII,
the gas metallicity ($[\ZnH]=\Zndla\pm0.05$) is much lower than that of the galaxy ($[$O/H$]=-0.35\pm0.1$~dex).
This comparison disfavor an outflow scenario because these tend to be metal rich \cite{TrippT_11a}.
Moreover, a bi-conical outflow should have a very wide cone opening angle ($>140$\degree) in order to intercept the quasar line-of-sight given the 
galaxy inclination of 55~$\deg$ (Fig.~S6).  Such a wide cone would result in a very large covering fraction, not supported by DLA host statistics, and would not be compatible with the current constraints on opening angles for outflows near star-forming galaxies at $z=0.1$ to 1 \cite{BordoloiR_11a,KacprzakG_12b,BoucheN_12a}.

The kinematics of the absorbing gas give us more clues about the nature of the gas. Thanks to 
our VLT/SINFONI-AO  observations giving us the orientation of the galaxy with respect to the line-of-sight, 
the gas kinematics show distinct features  (Fig.~1, fig. S5), and
these features can be put in the broader context of the host galaxy kinematics.
For instance,  
  the gas seen in absorption 26~kpc from the galaxy --- corresponding to 7 times the half-light radius $R_{1/2}=3.6$~kpc or one third of the virial size of the halo $R_{\rm vir}$ ---  is moving in the same direction as the galaxy rotation; that is, the gas appears to be co-rotating. Indeed,
the observed velocity field of this rotating galaxy (Fig.~1b) shows redshifted velocities in the direction of the quasar location, and in the  quasar spectrum 
 all of the  low-ionization ions \ZnII, \CrII, \FeII\ and \SiII\ 
tracing the cold gas also show redshifted velocities (Fig.~1c).

Thanks to the apparent quasar location on the sky being only 10\degree\ to 20\degree\ from the galaxy's projected major-axis, we can identify two distinct features in the low-ionisation lines.
The first feature comes from a quantitative comparison between the velocity field of the galaxy (Fig.~1b)  and the low-ionization kinematics (Fig.~1c): The velocities seen in absorption for  component 4 and seen in emission from the galaxy closely match one another.
This would indicate that a simple rotating disk
with circular orbits and a normal (flat) rotation curve can account for the bulk of the absorbing gas kinematics, but not all.
Thus,  components  1 to 3 which have a line-of-sight velocity less than the rotation speed  correspond to gas that is also co-rotating, but that is not rotationally supported  because these components do not have the required speed.  In other words, two groups of components (1 to 3;  4 to 6) are seen and one group is offset  by about 100\kms\ from the rotation pattern;
 hence this group cannot be gravitationally supported  and therefore should be flowing in. 
Each of these two groups contain about half of the total   \HI\ column  $\log\NH\simeq 20.6$ (fig.~S5) \cite{SOM}.

Thus, our SINFONI and UVES data show that the absorbing components have the broad characteristics of low-metallicity, co-planar, co-rotating accreting material similar to the expected features from numerical simulations \cite{StewartK_11b}. 
Moreover, the dust profile (fig.~S5D) determined from the data~\cite{SOM} indicates that  components 1 to 3  are less enriched than the main components (4 \& 5). Given that the amount of dust (or dust-to-gas ratio) correlates
with the metallicity \cite{VladiloG_06a},  the dust  profile also supports the scenario in which  
the quasar line-of-sight probes a mix of accreting material with that of an extended gaseous disk.

To gain further insights into the accreting gas,
we experimented with a simple geometrical toy model using a Monte Carlo approach to generate simulated absorption profiles \cite{SOM}. 
In the model, we distribute ``particles''  representing gas clouds or absorption components 
 where the cloud kinematics reflect an inflowing gas (radial for simplicity, see Fig.~2). 
Because the relative galaxy orientation is well determined
by the VLT/SINFONI data, the only free parameter is the inflow speed (fig.~S6).
The resulting   absorption profile simulated at the UVES resolution (Fig.~1D)
shows qualitative agreement with the data (Fig.~1C).

We do not expect such a model to be completely realistic nor to fit the data perfectly;
 there are many uncertainties such as the exact geometry of the accreting gas. However, the model can give us a framework to 
estimate the amount of accretion \cite{SOM}.  
The accreting material coming from the large-scale filamentary structure is presumed to
form a roughly co-planar structure   around the galaxy  \cite{StewartK_11b,ShenS_13a}.
Assuming  an azimuthal symmetry (see Fig.~2 and fig.~S6), 
the global mass flux rate through an  area of $2\pi\, b\,h_z$  is  thus
\begin{eqnarray}
\dot M_{\rm in}(b) %&=& 2\pi b\,h_z\,V_{\rm in}\,\rho(b) \nn\\
&\propto& 2\pi \,\NH\,b\, V_{\rm in}\, \cos(i), \nn \\
&=& 46\left(\frac{\mu}{1.6}\right)\left(\frac{\NH}{2\times10^{20}~{\rm cm}^{-2}}\right)\left(\frac{b}{26~{\rm kpc}}\right)\left(\frac{V_{\rm in}}{200~{\rm km/s}}\right)\left(\frac{\cos(i)}{0.57}\right) 
\mpy \label{eq:accr}
\end{eqnarray}
where $\mu$ is the mean molecular weight, $\NH$ is the gas column,  $b$ the impact parameter, $V_{\rm in}$ is the radial velocity component  
and  $i$ is the disk inclination.
This expression holds for any thin structure  of thickness $h_z$ and density $\rho$ because
we used the identity $\NH=\int{\rm d}z\rho(b)= \rho(b) h_z/\cos(i)$.

Using the constraints on $V_{\rm in}$ and on $\NH$ from our SINFONI and UVES data set,
the estimated mass flux $\dot M_{\rm in}$ in cold gas is around 30 to 60~\mpy 
 given the uncertainties in the column density and inflow speed.
This accretion rate  could be overestimated if the flow is very asymmetric and more realistic models with spiral orbits will lead to results that may differ by the cosine of the spiral opening angle, leading to (downward) corrections by factors of 2 to 3. Similarly, if we relax our assumption of a flat \HI\ profile, the gas column density must be higher for components 1 to 3 
 to account for both the dust profile and the \SiII\ column density, leading to an
upward correction to the accretion rate.

The range of accretion rates estimated from Eq.~1, 30 to 60~\mpy, is found to be close to the galaxy's SFR of $\sim$\sfr.
This is in agreement with  the simplest arguments for galaxy growth via  self-regulation  \cite{DuttonA_10a,BoucheN_10a}
and from numerical simulations \cite{DekelA_09a,FaucherG_11b}.
Furthermore, for this galaxy's halo mass,  $M_h\sim4\times10^{11}$~\msun\
(determined from its rotation curve),
  this value of $\dot M_{\rm in}$ corresponds to an accretion efficiency  $\epsilon$  of $\sim100$\%\
[where $\epsilon$ is defined as the ratio between the observed and maximum expected
baryonic accretion rate, namely $\epsilon\equiv\dot{M}_{\rm in}/(f_B\;\dot{M}_{\rm h})$, where
$f_B$ is the universal baryonic fraction and $\dot{M}_{\rm h}$ the halo growth rate \cite{GenelS_08a,McBrideM_09a}].

Our study shows the potential of the technique of using background quasars passing near
galaxies to further understand the process of gas accretion in galaxies. This technique is
complementary to other recent studies  %\cite{GiavaliscoM_11a,RubinK_12a,MartinC_12a,RibaudoJ_11a,KacprzakG_12a}.
\cite{GiavaliscoM_11a,RubinK_12a,RibaudoJ_11a}.
Our observations, which are merely consistent with cold accretion, provide key evidence important to consider against hydro-dynamical simulations.

%\bibliography{References and Notes}

% Following is a new environment, {scilastnote}, that's defined in the
% preamble and that allows authors to add a reference at the end of the
% list that's not signaled in the text; such references are used in
% *Science* for acknowledgments of funding, help, etc.

\section*{Acknowledgments}
 We thank the ESO Paranal staff for their continuous support and the SINFONI instrument team for their hard work which made SINFONI a very reliable instrument. 
NB   thanks  S. Lilly and R. Bordoloi  for stimulating discussions and S. Genel for a careful read of the manuscript.
We thank I. Schroetter for his assistance in making Figure 2.
We thank the reviewers for their thorough review, their comments and suggestions.
This research was based upon work supported in part by the National Science Foundation under grant No. 1066293 and the hospitality of the Aspen Center for Physics. 
It was partly supported by a Marie Curie International Outgoing
Fellowship (PIOF-GA-2009-236012) and by a Marie Curie International Career Integration
Grant (PCIG11-GA-2012-321702) within the 7th European Community Framework Program.
M.T.M thanks the Australian Research Council for a \textsl{QEII Fellowship} (DP0877998) and \textsl{Discovery Project} grant (DP130100568).
C.L.M. is supported by   NSF  grant AST-1109288.
 The data used in this paper are tabulated in the Supporting Online Material and archived at   \url{http://archive.eso.org} under program ID 383.A-0750 and 088.B-0715.
G.G.K. is an Australian Research Council Super Science Fellow.

\ifthenelse{\boolean{astroph}}{}{
\flushleft{\bf Supplementary Materials}\\
www.sciencemag.org \\
SOM text\\
Figure S1--S6\\
Table S1--S4 \\
References \\
}

\pagebreak

\noindent {\bf Fig. 1} Emission and Absorption Kinematics.
{\bf (a)} The color scale represents a narrow band image (rest-frame \Ha) with
 the continuum subtracted from our  AO-assisted VLT/SINFONI datacube.
The quasar HE2243$-$60 and the host $z\sim2.328$ galaxy are marked. 
The residuals from the continuum subtraction are visible both near the QSO and near the  position labeled R. 
The galaxy is  detected in \Ha\ with a maximum  SNR of $\sim5$--8 per pixel; no continuum emission
is detected. The beam has a full-width-at-half-maximum (FWHM) of $\sim0.25$ arc~sec. 
{\bf (b)} The fitted velocity field (extrapolated over the entire field) 
is shown along with the flux contours. 
The kinematic parameters were determined using our 3D analysis \cite{SOM}.
The dotted line shows the kinematic major axis.
At the quasar location (solid circle),  the rotation speed is expected to be $\sim160$ to 180~\kms.
{\bf (c)} The absorption profiles from the VLT/UVES spectra showing the line-of-sight velocity
of the various absorption components in
the low-ionization ion  (\ZnII, \CrII, \NiII,  \FeII and \SiII) where  $v=0$~\kms\ corresponds to 
the galaxy redshift.
The main component (component 4)  contains half of the \SiII\ column density and appears to contain more dust \cite{SOM}.
This   component has a line-of-sight velocity   $v=180$~\kms\
consistent with that of the galaxy velocity field shown in (a).
{\bf (d)} A simulated line-of-sight profile for the geometry
of this QSO-galaxy pair from a toy model that includes an inflow component (black)
in addition to a component determined from the extended galaxy velocity field (red).
A schematic representation of the model is shown in Fig.~2 and a more detailed representation
is presented in fig.~S6. 

\ifthenelse{\boolean{astroph} }{
\begin{figure}
\centering
\includegraphics[width=15cm]{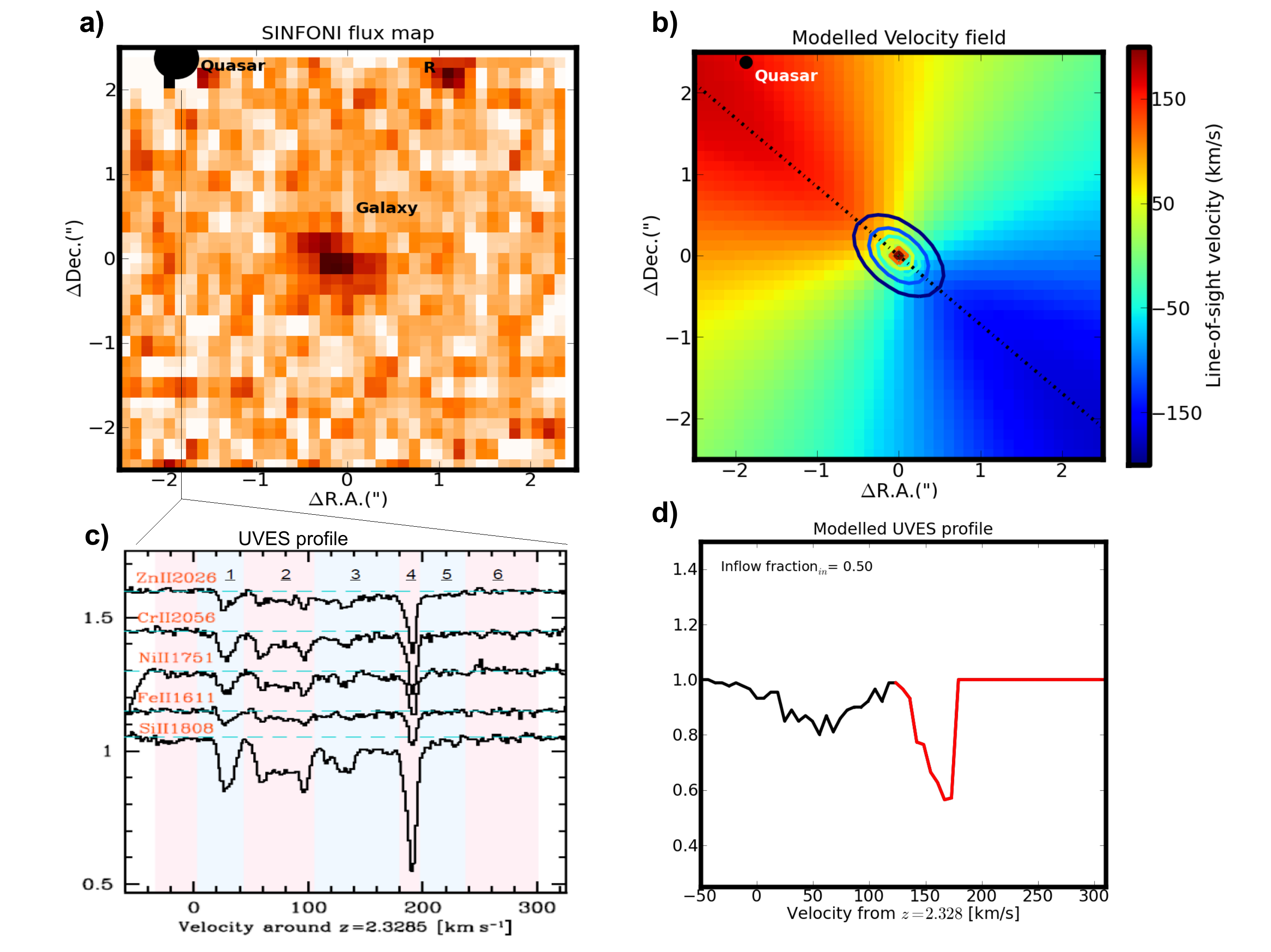}
\end{figure}
}{}

\pagebreak

\noindent{\bf Fig. 2} Schematic diagram of the simplified model.
The star-forming galaxy is experiencing a super-nova driven outflow (red)
and receiving gas with low angular momentum  from the inter-galactic medium (blue).
The opening angle of the bi-polar outflow is assumed to be approximately 60\degree,
as found in low-redshift observations \cite{BordoloiR_11a,BoucheN_12a,KacprzakG_12b}.
The accreting cold gas is expected to dissipate its angular momentum and migrate
towards the galaxy, delivering fuel for the galaxy and forming an extended gaseous co-planar structure. 
The quasar (yellow) line of sight is represented by the star and is about $\alpha=20$\degree\ from the major-axis leading to distinct kinematic features seen in the absorption spectra (Fig~1d; fig.~S6).  
For the observed geometric configuration, the line-of-sight is intercepting the accreting material  but not the outflowing material.  

\ifthenelse{\boolean{astroph} }{
\begin{figure}
\centering
\includegraphics[width=18cm]{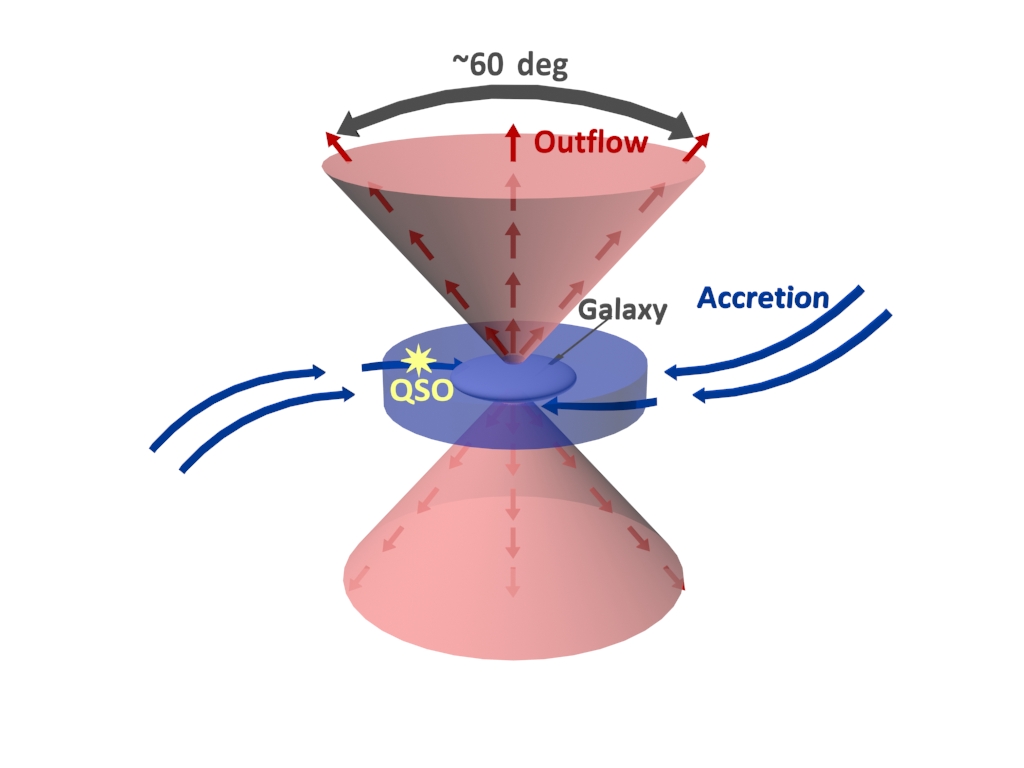}
\end{figure}
}{}

\clearpage

\begin{table}
 \centering
 \begin{minipage}{140mm}
  \caption{Physical properties of the galaxy\label{table:summary}}
  \begin{tabular}{lll}
  \hline
Quasar Impact Parameter $(b)$ 		& 3.1$"$ or 26~kpc \\
Major-Axis position angle			    & 62$\pm$5~\degree\\
Quasar position angle				    & 42$\pm$1~\degree\\
Redshift $z_{\Ha}$			& \zHa$\pm$0.0001  \\ 
Half-Light Radius	($R_{1/2}$)		& 3.6$\pm0.3$~kpc\\
Maximum Circular Velocity ($V_{\rm max}$) 	& 150$\pm$15~\kms \\
Inclination	($i$)		 & 55$\pm$1 ~\degree\\
Dispersion 	Velocity ($\sigma_i$)		    & 158$\pm5$~\kms\\
Halo Mass	($M_h$)				& $\sim4\times10^{11}$~\msun\\
Virial Radius ($R_{\rm vir}$)	    & $\sim90$~kpc\\
Dynamical Mass~\footnotemark[1]  ($M_{\rm dyn}$)			& $2.0\times10^{10}$~\msun \\ 
Interstellar Gas Mass~\footnotemark[2]	($M_g$)				& $1.5\times10^{10}$~\msun\\
Gas fraction	($f_g$)			& $\sim$0.75 \\Metallicity	($\logZ$)		& $-0.35\pm$0.1\\
SFR$_{\Ha}$~\footnotemark[3]					& 18$\pm2$~\mpy	&\\
Dust corrected SFR$_{\Ha}$	& \sfr	&\\
Gas Accretion Rate	($\dot M_{\rm in}$)			& $\sim$45~\mpy\\
\end{tabular}
\end{minipage}
\end{table}
\footnotetext[1]{Within the half-light radius $r_{1/2}$ assuming $V_c^2=G\;M/r_{1/2}$.}
\footnotetext[2]{Within the half-light radius $r_{1/2}$ determined from the \Ha\ surface density by inverting the star-formation law.}
\footnotetext[3]{Integrated SFR~\cite{SOM}.}
\clearpage

\clearpage

%%%%%%%%%%%%%%%%%%%%%%%%%%%%% SOM

\startsupplement

{\Huge Supplementary Online Materials for}\\ \\

{\center \Large{Signatures of Cool Gas Fueling a Star-Forming  Galaxy}} 
{\center \Large{at Redshift 2.3}}
\\
N. Bouch\'e$^{\ast}$, M. T. Murphy, G. G. Kacprzak, C. P\'eroux,\\
T. Contini, C. L. Martin, M. Dessauges-Zavadsky\\
 
\normalsize{$^\ast$To whom correspondence should be addressed. E-mail: nicolas.bouche@irap.omp.eu.}

\date{} % Activate to display a given date or no date (if empty),
         % otherwise the current date is printed 

\maketitle
\thispagestyle{empty}

{\bf This pdf includes:}\\
\\
Supporting Online Material:\\
Figs. S1--S6\\
Tables S1--S4\\
References 31--73\\

\ifthenelse{\boolean{astroph}}{ 
    \setcounter{page}{1}
}{
    \setcounter{page}{0}
}

%\section{Supporting Online Materials}

In this Supporting Online Materials  we provide additional details on
the data sets and the data analysis regarding the accretion of cool gas near a
$z\sim2.3$ star-forming galaxy.
In section \S~\ref{section:sinfo}, we describe the VLT/SINFONI observations and we present  our 3D kinematic analysis.
In section \S~\ref{section:uves}, we describe the  VLT/UVES observations and present the absorbing 
gas kinematics seen along the quasar line-of-sight.
In section \S~\ref{section:model}, we present more information regarding the toy model used in the main text.
Throughout, we use   the following cosmological parameters
$h=0.7$, $\Omega_M=0.3$, and $\Omega_\Lambda=0.7$.

\section{Galaxy emission properties}
\label{section:sinfo}

\subsection{SINFONI/VLT Integral Field Spectroscopy}

The star-forming galaxy    is associated with a damped \Ly\ absorber  (DLA) with a neutral hydrogen column density\NHI\ of $\log [\NHI/\cmsq] = 20.62\pm0.05$ and was found 26~kpc (3.1\arcsec) away  from the quasar HE 2243$-$60 in
seeing-limited VLT/SINFONI observations~\cite{BoucheN_11a}. 
In this study, we use recent VLT/SINFONI observations, obtained in service mode between 2009 and 2011 (see Table~\ref{table:observations}). In order  to ensure the highest spatial resolution possible the observations were taken using Adaptive Optics 
(AO) with the quasar as a Natural Guide Star (NGS). 
The field was observed in all three gratings ($J$, $H$, and $K$),
covering all  the nebular lines \OII,  \OIII$+$\Hb\ and \Ha$+$\NII.
We used the 0.125\arcsec\ pixel scale in all three
gratings  to maximize the surface-brightness sensitivity. 
The total integration times were 12ks, 12ks and 10.8ks in $K$, $H$, and $J$ respectively. 
The full-width-at-half-max (FWHM) of the point spread functions (PSF) are 0.25\arcsec, 0.4\arcsec\ and 0.65\arcsec in $K$, $H$ and $J$ respectively.  The data reduction was performed  
 using the MPE SINFONI pipeline  \cite{SchreiberJ_04a,AbuterR_06a} complemented with additional routines designed to optimize the subtraction of
the OH sky lines  \cite{DaviesR_06a}. 
In addition, we applied the heliocentric correction
to the individual cubes before combining them to a single cube.
The flux calibration of the data was performed using the 2MASS
broadband magnitudes of standard stars ($O$ or $B$ stars), which yields a calibration  accurate to
$\sim15$\%.  The full description of the reduction steps is described elsewhere \cite{ForsterSchreiberN_06a,ForsterSchreiberN_09a,BoucheN_07a,BoucheN_11a}.

\subsection{Flux, Metallicity and Size measurements}
\label{section:flux}

\begin{figure}
\centering
\subfigure[]{
\includegraphics[width=6.5cm,height=4cm]{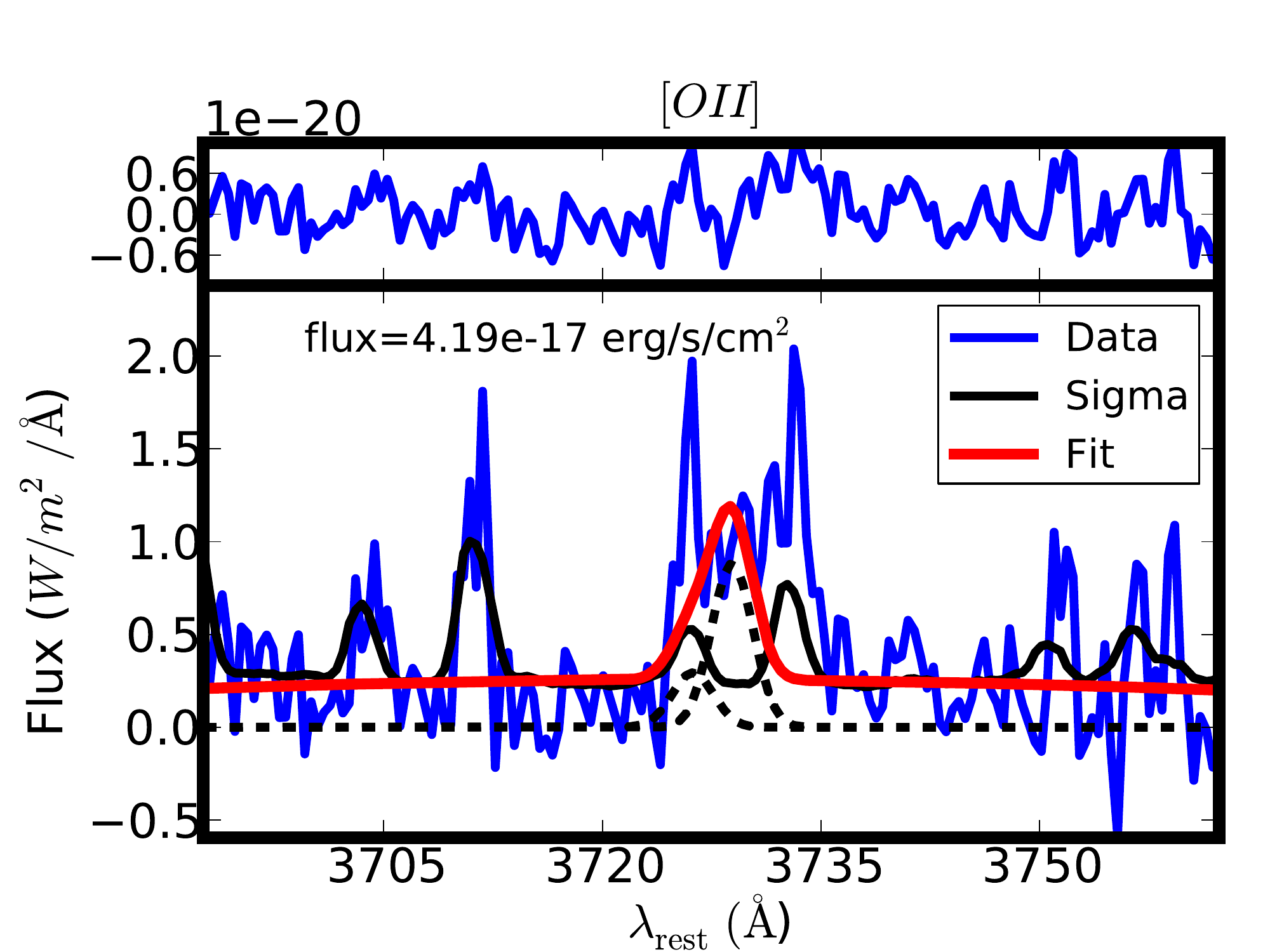}
}
\subfigure[]{
\includegraphics[width=6.5cm,height=4cm]{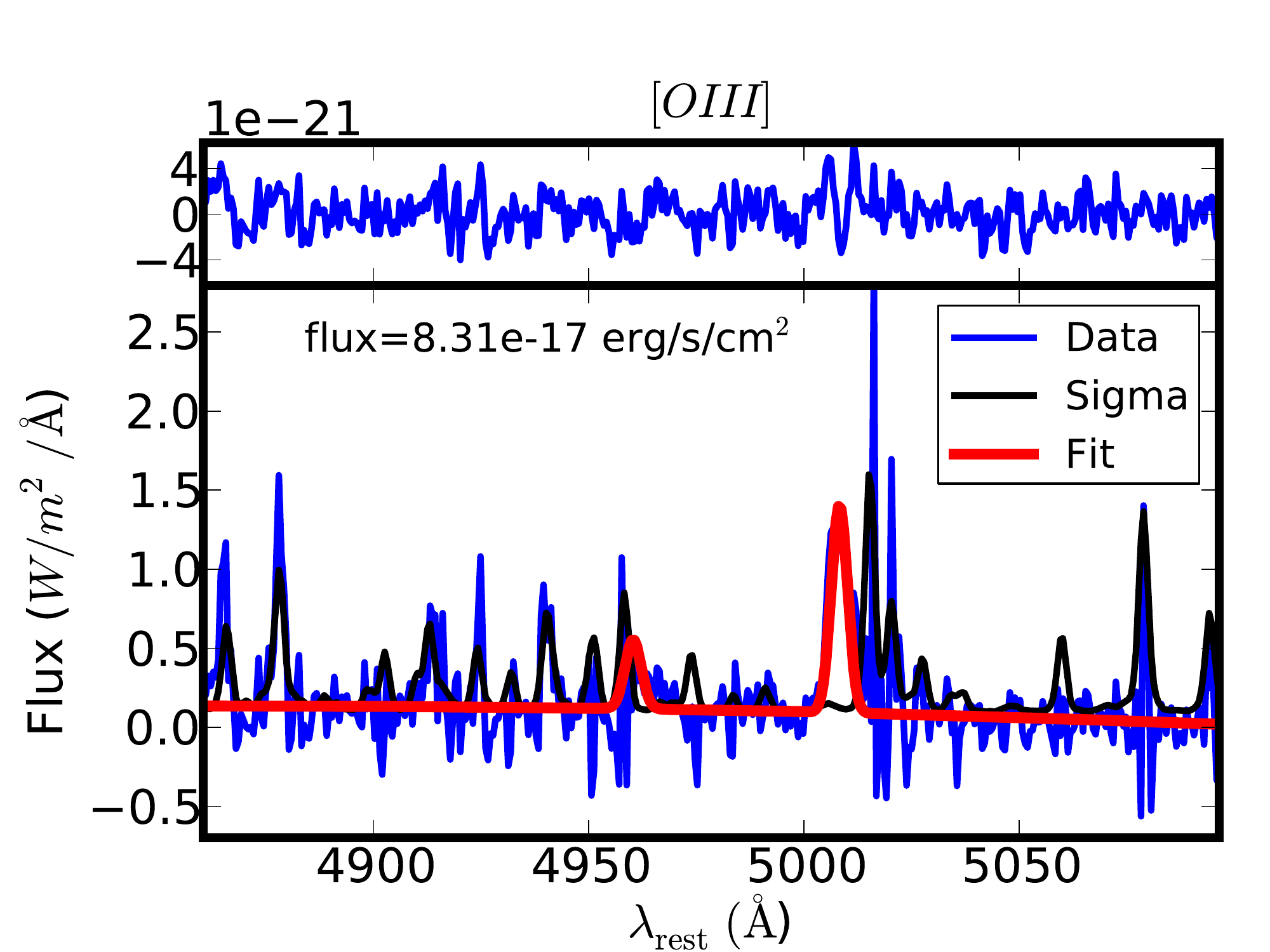}
}
\subfigure[]{
\includegraphics[width=6.5cm,height=4cm]{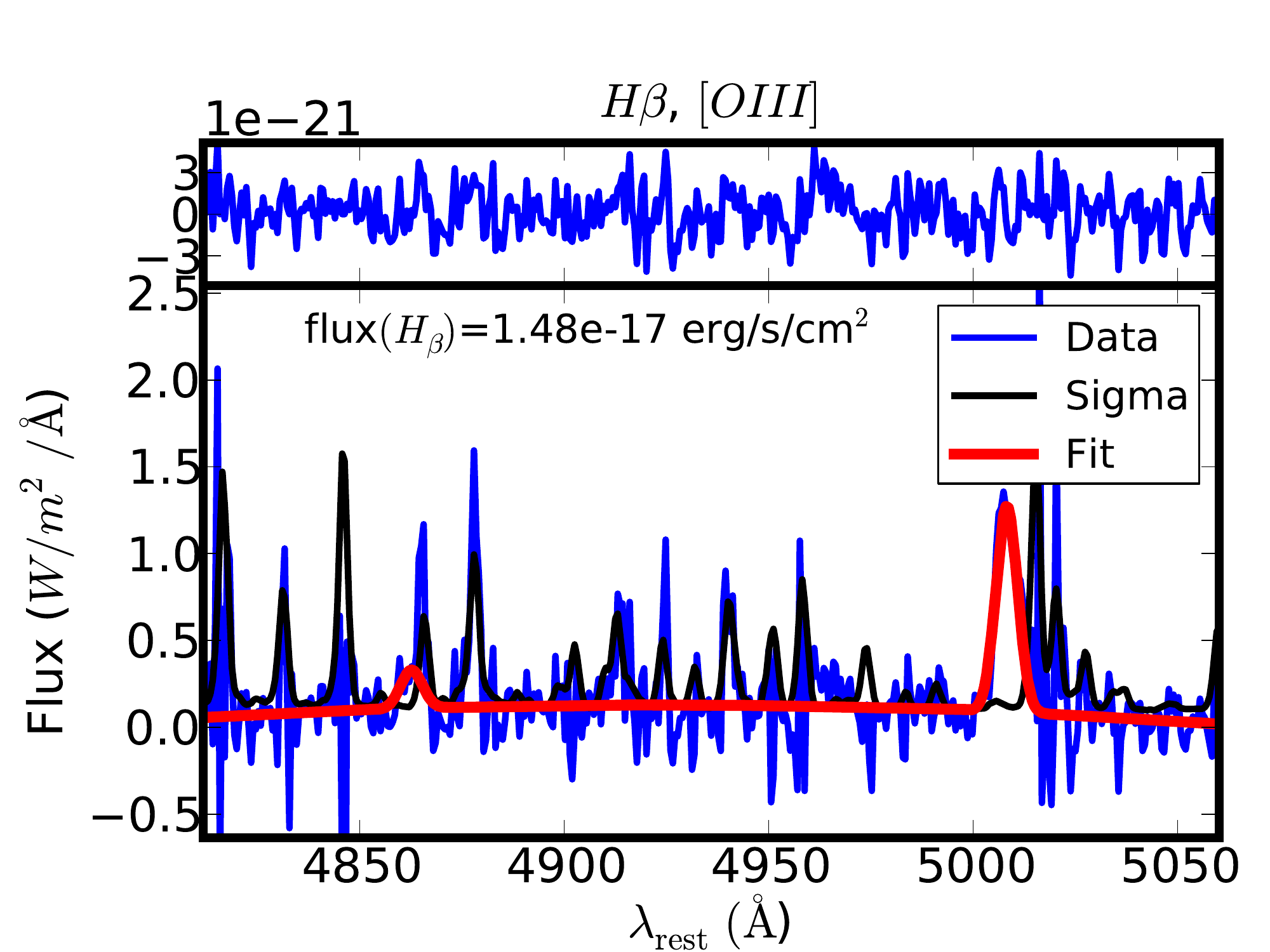}
}
\subfigure[]{
\includegraphics[width=6.5cm,height=4cm]{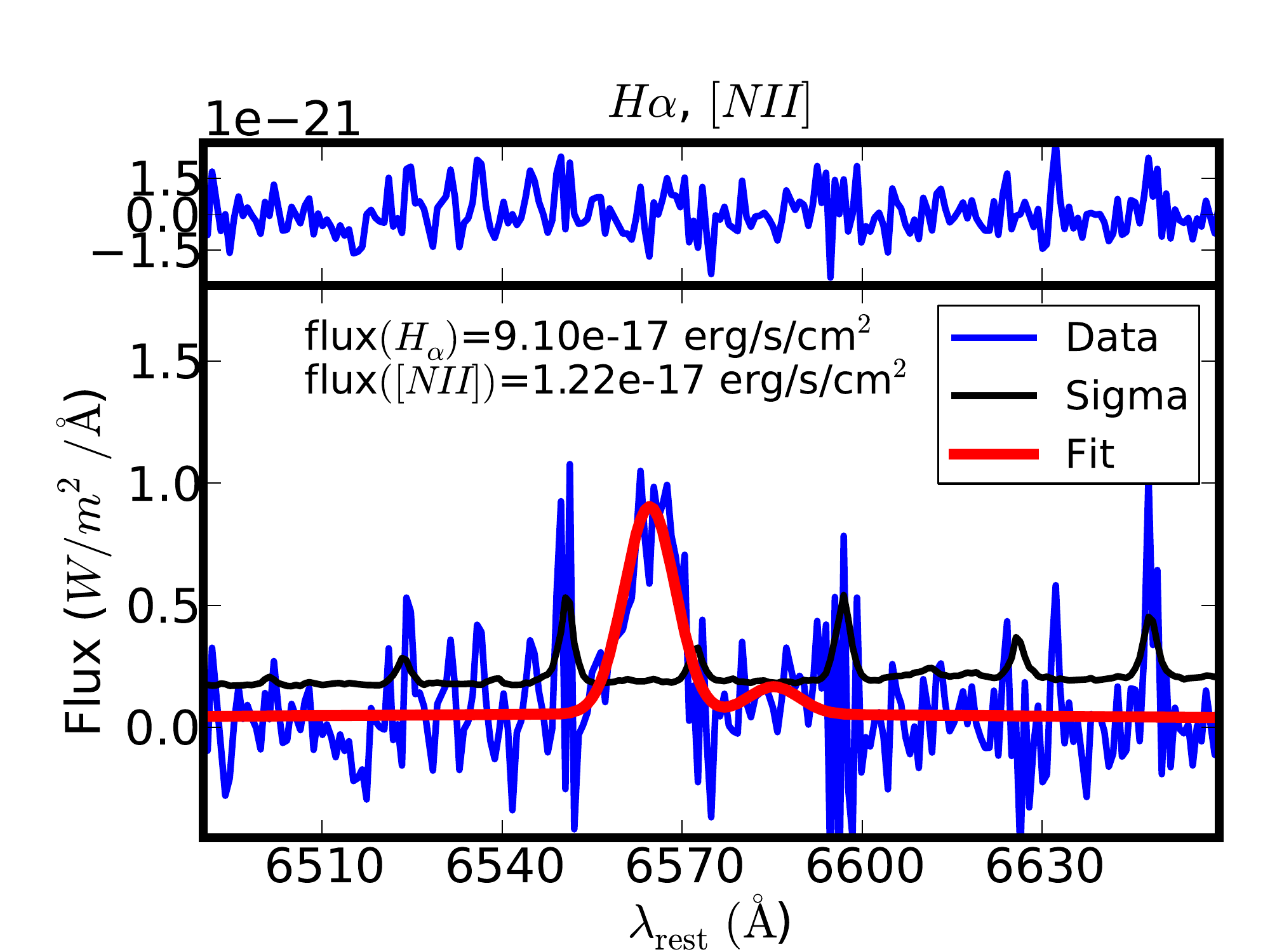}
}
\caption{{ The SINFONI 1D-spectra extracted in a 1.25\arcsec\ radius aperture.}
{\bf (a)} $J$-band \OII\ spectra.
{\bf (b)} $H$-band \OIII\ spectra.
{\bf (c)} $H$-band \Hb\ spectra.
{\bf (d)} $K$-band \Ha+\NII\ spectra.
In each panel, the data is represented by the blue lines, the error vector by the black line,
and the fitted emission line(s) is shown by the thick red line.
For each spectrum, the top sub-panel shows the residuals.
 The total flux determined from the fit is labeled.
The two components
of the \OIII$\lambda\lambda4959,5007$ and the \OII$\lambda\lambda3725,3727$ doublets
 are fitted simultaneously. The two components of the \OII\ doublet are shown with the dashed lines.
For \Hb, a joint fit is performed jointly with \OIII$\lambda5007$.
For \NII$\lambda6584$, a joint fit is performed with \Ha.
Note the presence of OH line in the \OII\ and \OIII\ lines. 
\label{fig:spectra}}
\end{figure}

Figure~\ref{fig:spectra} shows the one-dimentional 
spectra extracted in a 1.25\arcsec\ radius aperture for each of the nebular lines.
We fitted each of the emission lines with Gaussian line profiles. 
The two components of the \OIII\ and \OII\ doublets are fitted simultaneously. 
The \NII$\lambda$6584 emission line is fitted jointly
with the \Ha\ line (with the reshift and dispersion joined), and  the \NII$\lambda$6584 flux should be treated as an upper limit.
The \NII\ flux is  $<1.2\times10^{-17}$~\flux. 
Similarly, the \Hb\ flux is determined from a joint \OIII$+$\Hb\ fit where the centroid and line width (FWHM) are tied.
The flux measurements are summarized in Table~\ref{table:param}.
From the \Ha\ flux of $9.1\pm0.3\times10^{-17}$~\flux, we 
infer an observed SFR of $\sim18\pm2$~\mpy for a Chabrier   Initial Mass Function (IMF) \cite{ChabrierG_03a}, and a
dust-corrected intrinsic SFR$_0$ of $\sim30\pm5$\mpy,
using $E(B-V)\simeq 0.25$ estimated below
and a Calzetti  extinction curve~\cite{CalzettiD_00a}.
The \OIII\ flux is $8.1\pm0.4\times 10^{-17}$~\flux.
We find that the \Hb\ flux is $1.7\pm 0.4\times 10^{-17}$~\flux.
The \OII\ flux is $3.8\pm0.4\times10^{-17}$~\flux. 
See Table~\ref{table:flux} for the summary of the flux measurements.

\begin{figure}
\centering
\includegraphics[width=16cm]{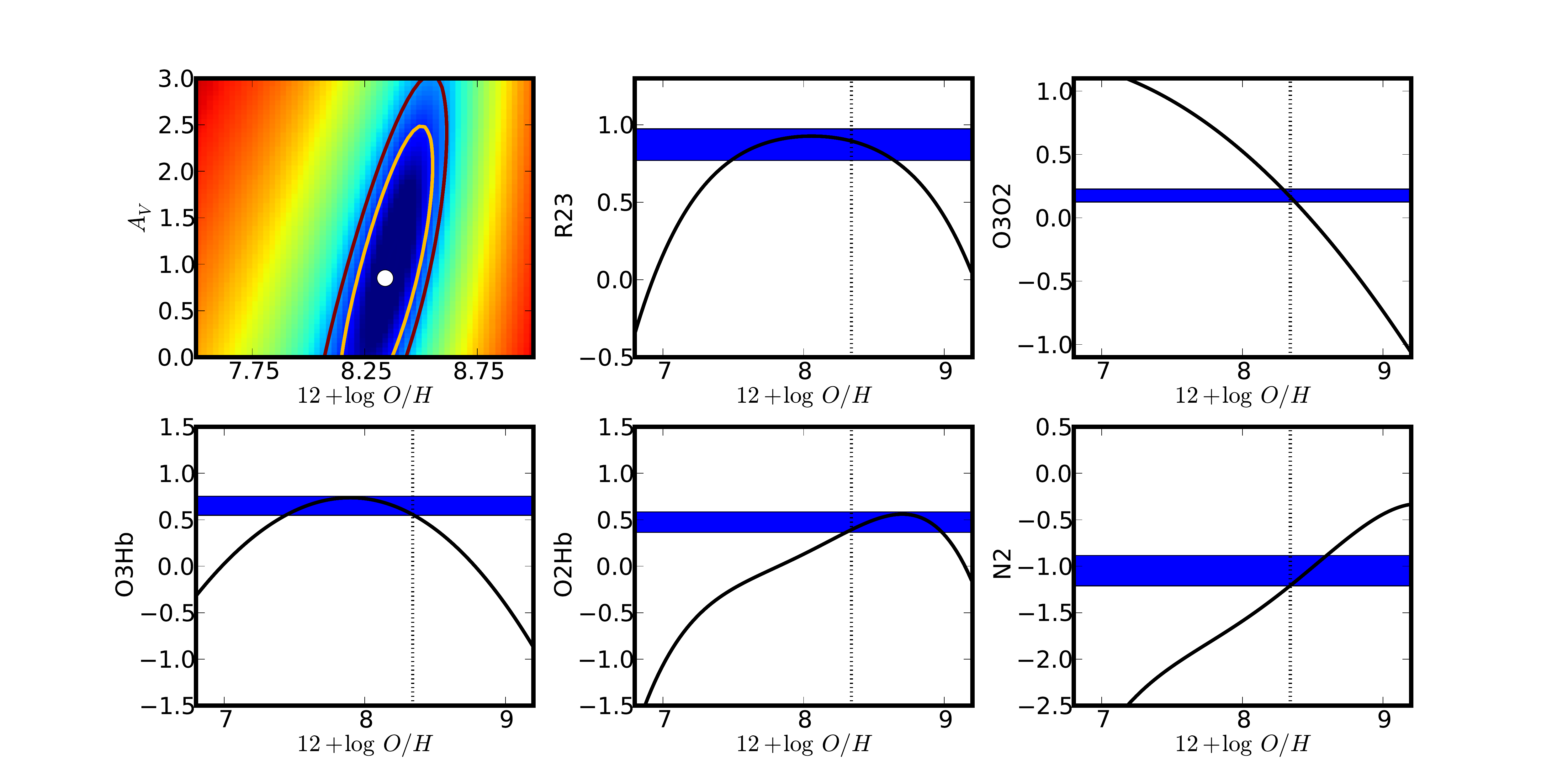}
\caption{{ Galaxy Metallicity and Extinction.}
Using five metallicity indicators (R23, O3O2, O3Hb, O2Hb, N2),
we fit simultaneously the metallicity and the exctinction.
The color map shows the $\chi^2$ surface, with the $1\sigma$ and $2\sigma$ levels shown as contours.
The fitted metallicity is found to be $12+\log \OH=8.35\pm0.10$ and the 
fitted extinction is $E(B-V)\sim0.25$, corresponding to  $A_V\sim0.8$ and $A_{\Ha}=0.65$. \label{fig:he2243}}
\label{fig:maiolino}
\end{figure}

With our \OII, \OIII$+$\Hb, and \NII$+$\Ha\ flux measurements we can determine the metallicity
of the host using the five common metallicity indicators (R23, O3O2, O3Hb, O2Hb, N2) \cite{KewleyL_02a}.
Because the \NII\ line is only hinted in the data (see Figure~\ref{fig:spectra}) and the R23 indicator is double-valued, 
the tightest constraint comes from O3O2.  We use a simultaneous fit  to the five indicators 
 to determine the metallicity $Z$ and  the amount of extinction $A_V$ \cite{MaiolinoR_08a}.
Figure~\ref{fig:maiolino} shows that the metallicity of the host towards HE 2243$-$60 is 
\begin{equation}
12+\log\,\frac{O}{H}= 8.34\pm0.10 \label{eq:Z}
\end{equation}
corresponding to a metallicity of $Z=-0.35\pm0.1$ using solar abundance $Z_\odot=8.69$
\cite{AllendePC_01a}, which is  typical for galaxies at $z\gtrsim2$ \cite{CresciG_10a,MannucciF_10a}.
In other words, the galaxy seems to lie on the mass--metallicity sequence 
and  we find no evidence for a metallicity gradient from the emission line galaxy data.
From our global fit to the line fluxes, the extinction is found to be  $E(B-V)\sim0.25$, corresponding to $\sim$0.65~mag of extinction ($A_{\Ha}$)
and $A_{V}\sim0.8$ mag.

We collapsed the data cube in 2D flux maps and found that the \Ha\ flux maps are well described by a 2D exponential profile
(Figure~\ref{fig:profile}). We find that the galaxy has
 a disk scale length of $R_{d}=0.27\pm0.02\arcsec$ corresponding to $2.3\pm0.2$~kpc.
From the flux curve-of-growth $f(<r)$, the \Ha\ half-light radius is $R_{1/2}=0.44\pm0.03$\arcsec\ or 3.6$\pm0.3$~kpc.
The \OIII\ surface brightness profile gives a value identical within the errors.  
The \OII\ surface brightness profile is poorly constrained because of the low \OII\ flux, significant OH contamination in the wings of \OII, and of the lower spatial resolution (0.65\arcsec\ in the $J$-band).

\subsection{Emission Kinematics and 3D Modeling}
\label{section:kine}

We extracted the velocity field and dispersion map 
using the code LINEFIT developed  specifically for SINFONI
applications \cite{ForsterSchreiberN_06a,ForsterSchreiberN_09a,DaviesR_11a}.   
The main key features of LINEFIT are that  
the spectral resolution is explicitly taken into account by
convolving the assumed intrinsic emission line profile and a
template line shape for the effective instrumental resolution
before performing the fits.
Figure~\ref{fig:linemap}(left) shows the flux map, 2D velocity field   and dispersion maps
 for the \Ha\ line (bottom row) and the \OIII\ line (top row).
Both the \Ha\ and the \OIII\ line trace the kinematics of the galaxy.

Contrary to traditional methods to construct 2D velocity and dispersion maps
\cite{PuechM_08a,CresciG_09a,WrightS_09a,EpinatB_09a,EpinatB_10a},
we extracted the kinematic properties of the host galaxy directly from the data cube
 using a 3D fitting algorithm,  i.e. without collapsing the data in 2D sub-products. 
Essentially, for a given flux profile and a kinematic
profile, we search for the best parameters that describe the data
using a Markov Chain Monte Carlo (MCMC) Bayesian approach.
The algorithm compares the data with the model convolved with the Point Spread Function (PSF) and spectral Line Spread Function (LSF) and returns both the model parameter and the intrinsic 3D model, i.e. unconvolved model.
This method has several advantages, such as the dynamical center
is constrained by the data (and thus need not to be fixed),
the creation of 2D velocity maps is not necessary, and it provides
us with a `deconvolved' model.

The parametric model used in the MCMC algorithm includes  a  geometrically thick disk with an
exponential flux profile $f(r)\propto \exp(-r/R_d)$.
The disk intrinsic thickness $h_z$ is taken such that the edge-on axis ratio is $q\equiv h_z/R_{1/2}=0.2$, 
corresponding to $\sim1$~kpc, typical of high-redshift edge-on/chain galaxies \cite{ElmegreenB_06a}, and
the flux profile  in the direction perpendicular to the orbital plane is modeled  as $f(z)\propto\exp(-0.5\,z^2/h_z^2)$.
For the disk kinematics, we used parametric description of the rotation curve $v(r)$. 
We use  either (1) an arctangent curve \cite{PuechM_08a},  $v(r)\equiv 2/\pi V_{\rm max}\arctan(r/r_p)$, which has two parameters $V_{\rm max}$ the asymptotic velocity and $r_p$ the inner slope,
or (2)   an exponentially rising profile \cite{FengJ_11a}, $v(r)=V_{\rm max}(1-\exp(-r/r_p))$.
We will refer to these two types of rotation curve as `arctan' or `Exp', respectively.

The total (line-of-sight) velocity dispersion $\sigma_{\rm tot}$ includes three  terms (added in quadrature).
It includes (i) the local isotropic velocity dispersion $\sigma_d$ driven by the disk self-gravity is $\sigma_d(r)/h_z=V(r)/r$ 
for a dynamically hot thick disk \cite{GenzelR_08a, BinneyJ_08a,DaviesR_11a},
(ii) a mixing term,  $\sigma_m$, arising from mixing the velocities along the line-of-sight for a geometrically thick  disk,
and (iii)  an intrinsic  dispersion ($\sigma_i$)  ---often dominant---  
to account for the fact that high-redshift disks are dynamically hotter than the self-gravity expectation.
In $z>1$ disks,  $\sigma_i$ is  often observed to be $\simeq50$--80\kms\ \cite{LawD_07a,GenzelR_08a,WrightS_09a,CresciG_09a,ForsterSchreiberN_09a,EpinatB_10a,EpinatB_12a,WisnioskiE_11a,LawD_12a}.

Because the $\chi^2$ hyper-surface is very flat,  simple $\chi^2$ optimization tends to not converge as it is very susceptible to local minima. Hence, we use a Monte-Carlo Markov-Chain optimization algorithm  to optimize the parameters using flat priors on bound intervals for each of the parameters. 
The values (errors) of the fit parameters  are taken from the median (standard deviation)
 of the last 1000 iterations  of 5000 since convergence is usually achieved in the first 500 iterations.
We tested the algorithm extensively on about 200 data cubes
generated with known parameters covering a range of fluxes and 
seeing conditions.  We find that the  algorithm reproduces the input parameters within 10\%\ for our observing conditions (integration time and PSF).

\begin{figure*}
\centering
\includegraphics[width=14cm,height=5cm]{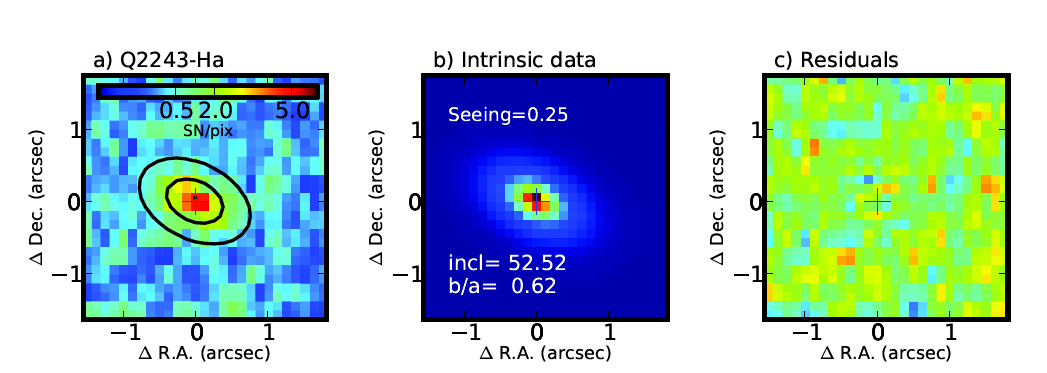}
\includegraphics[width=9cm,height=5cm]{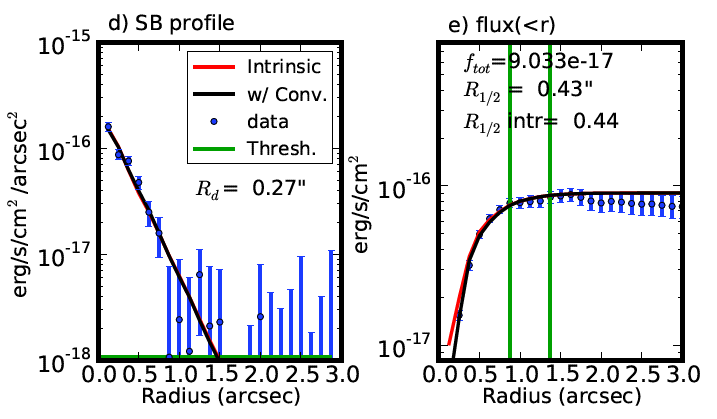} 
\caption{{ Flux Map and Surface Brightness Profile}.
From the observed \Ha\ flux map shown in {\bf a}, we fitted an exponential profile  convolved
with a Gaussian seeing  of FWHM=0.25\arcsec\
{\bf b}, whose residuals are shown in {\bf c}.
In {\bf a}, the color scale shows the SNR per pixel and the flux contours are shown.
The surface-brightness profile summed over annuli of increasing radius is shown in {\bf d}
and the corresponding growth curve $f(<r)$ profile is shown in {\bf e}.
In (d) and (e), the intrinsic (convolved) profiles are shown in red (black) respectively. 
The data is represented by the filled circles with error bars.
The errorbars are determined from bootstrap Monte Carlo resampling.
The horizontal green line in (d) shows the threshold where the SNR reaches unity.
The annuli where SNR reaches unity defines the  region 
from where we estimate the total flux ($f_{\rm tot}$) shown as the
vertical green lines in panel (e).}
\label{fig:profile}
\end{figure*}

\begin{figure*}

\centering
\subfigure[]{
\includegraphics[height=7.5cm,angle=0]{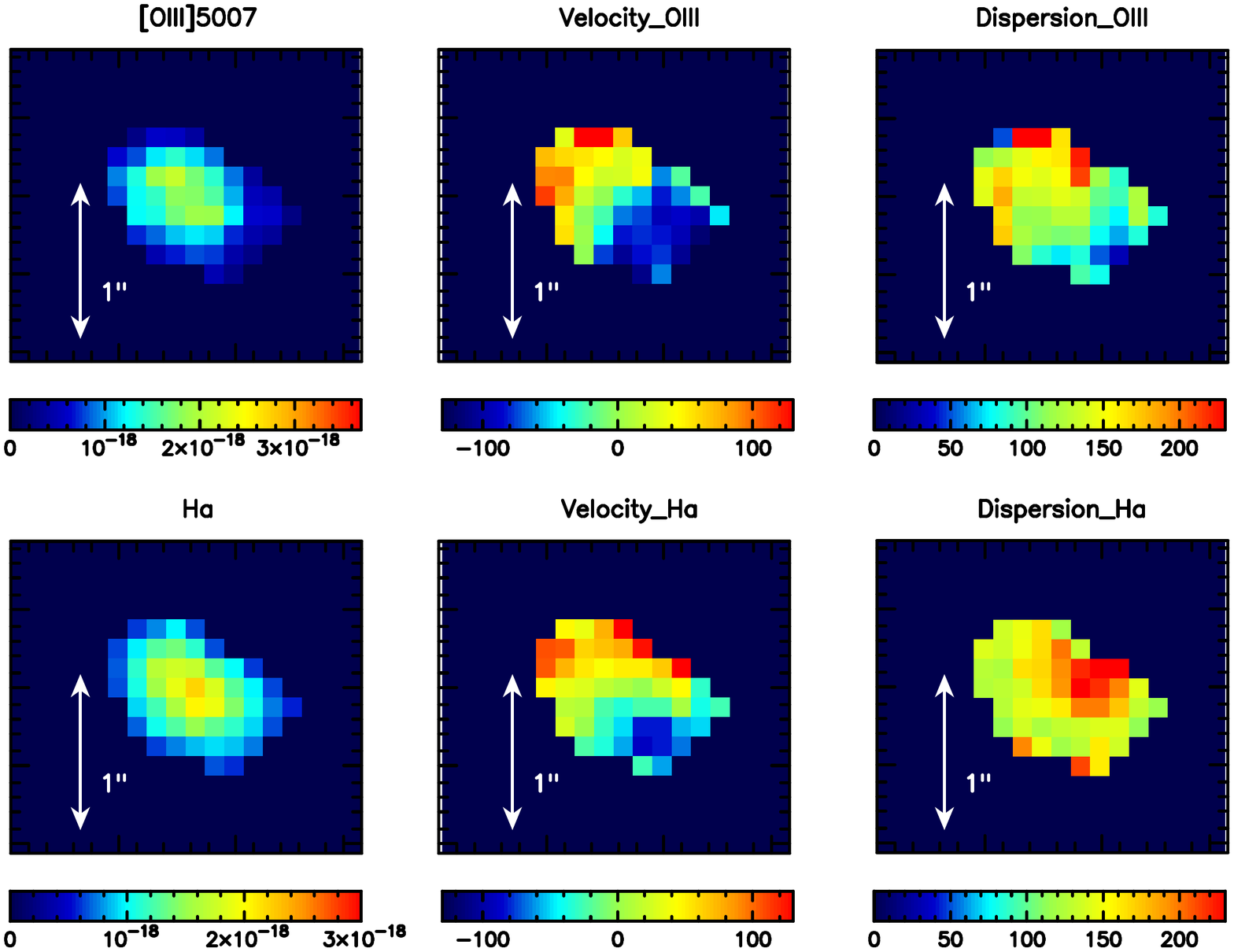}
}
\subfigure[]{
\includegraphics[width=4.4cm,height=7.5cm]{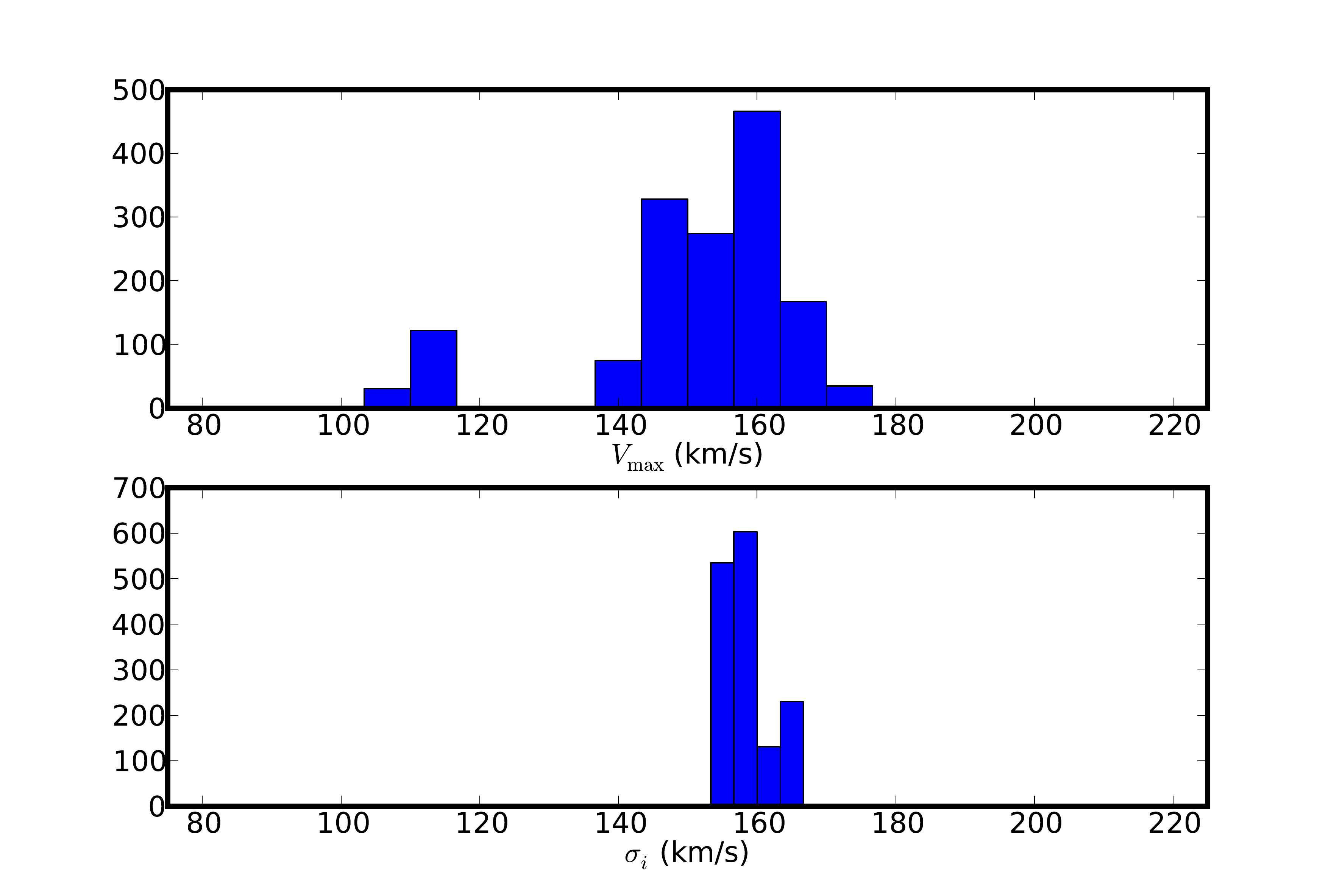}
}
\caption{{ Kinematic analysis of the $z=2.328$ galaxy}.\label{fig:linemap}
For the galaxy towards HE2243$-$60, {\bf a} shows the flux map (\flux), velocity field (\kms) and dispersion map (\kms)
are shown on the left, middle, right panels. The \Ha\ (\OIII)  maps are shown in the bottom (top) row,  where the FWHM of the PSF is 0.25\arcsec (0.4\arcsec), corresponding to $\sim$ 2~kpc (3.3~kpc), respectively. 
In {\bf b}, from our Markov chains on our 3D model parameters, we show distributions allowed by the data 
on the (intrinsic) maximum circular velocity $V_{\rm max}$ and the intrinsic dispersion ($\sigma_i$) parameters 
(using the arctangent model for the rotation curve).
}
\end{figure*}

In total, the  model has 10 free parameters to be determined from the data.
The 10 parameters are the $x_o$, $y_o$, $z_o$ positions, 
$R_d$ exponential scale length, the total flux, inclination $i$, position angle PA, $r_v$ and $V_{\rm max}$
the rotation curve parameters and $\sigma_z$ the intrinsic dispersion. 
The parameters determined from \Ha~\footnote{We did not perform the 3D analysis on the  \OIII\ data because it is affected by an OH-line on the red side (see Figure~\ref{fig:spectra}).} for our galaxy towards HE 2243-60  are listed in Table~\ref{table:kinematics}. For instance, 
the galaxy PA of $\sim 55\pm1~\degree$ is within 20 degrees of the quasar PA
which is 42~$\degree$, implying that the QSO line-of-sight is aligned within $\sim20~\degree$
from the host  major axis, which implies that the QSO intercepts the disk plane at a very favorable angle
to study the connection between the absorption and emission kinematics. 
The morphological parameters (size, $i$, PA) derived from this global 3D fitting technique on the \Ha\ data cube agree well with those derived from  the 2D-\Ha\ flux profile (Figure~\ref{fig:profile}).

In Figure~\ref{fig:linemap}a, we show the 2D velocity and dispersion maps for \Ha\ and \OIII.  In Figure~\ref{fig:linemap}b, we show
the distributions obtained from the MCMC algorithm for two of the model parameters. The maximum circular velocity is found to be $V_{\rm max}=145\pm16$~\kms, or $180\pm32$~\kms\ for the rising exponential or arctangent rotation curve, respectively. 
The distribution of $V_{\rm max}$ values allowed by the data given our model
is shown in the top panel in Figure~\ref{fig:linemap}b.  This corresponds to a dynamical mass of 
$\log (M_{\rm dyn}[\msun])\;\simeq 10.5$ within $2\times R_{1/2}$.
 Assuming that the maximum velocity $V_{\rm max}$ is a good proxy for the halo circular velocity $V_c(\sim V_{\rm vir})$, 
the total halo mass $M_{\rm vir}$  is found to be $\log( M_{\rm vir}[\msun])\sim$11.6.

The  intrinsic dispersion $\sigma_i$ required for this galaxy  is large, with an average of $\sigma_{i}=157\pm5$~\kms.
Interestingly, the 
2D dispersion map shows a peak at $\sim250$~\kms\ in
 a location off-centered from the dynamical center, roughly on the minor axis
 as shown in Figure~\ref{fig:linemap}a.  
The range of $\sigma_{i}$ values allowed by the data  shown
in the bottom panel in Figure~\ref{fig:linemap}b is large
compared to other rotationally supported galaxies at those   redshifts~\cite{ForsterSchreiberN_09a,CresciG_09a,WisnioskiE_11a,LawD_12a}, which have 
$\sigma_i\simeq50$--100~\kms.
 Such a large dispersion could be the signature of a significant amount  of shocked gas 
produced in a super-novae driven outflow \cite{NewmanS_12a,SotoK_12b} which can
leave an additional blue-shifted broad component with FWHM$\sim500$ \kms.
The \OI$\lambda$6300/\Ha\ line ratio is perhaps the most sensitive diagnostic for shocks \cite{DopitaM_95a} and
while we do not detect the \OI$\lambda$6300 line directly, we find an  upper limit of $f_{\OI}\lesssim 1.0\times10^{-17}$~\flux, leading to
a flux ratio $\OI/\Ha=0.1$, typical for star-forming galaxies.
Because, the shock velocities required for having a $\OI/\Ha$ ratio departing from the locus of star-forming galaxies 
is $\gtrsim250$--300~\kms\ \cite{DopitaM_95a,SotoK_12b}, the $\OI/\Ha$ ratio is consistent with both
star-formation and a mild shock with a shock speed $v_{\rm sh}$ which is similar to  our dispersion of $\sim150$ \kms.

\section{Absorbing gas properties}
\label{section:uves}

\subsection{High-resolution spectroscopy of the quasar HE 2243$-$60}

We  combined  observations of archival VLT/UVES data of the quasar
HE 2243$-$60 ($m_V=18.3$) from several programs (ESO Program ID 065.A-0411, 067.A-0567 \&\ 074.A-0201)
\cite{LopezS_02aa,NoterdaemeP_07a,FoxA_07b} 
 leading to a spectrum with a very high signal-to-noise  ratio (SNR).
The SNR   is $\sim$110 per 2~\kms\ pixel at 4000~{\AA} and 6800~{\AA} near
the Ly$\alpha$ and the \ZnII/\CrII\ absorption lines, respectively. 
The  SNR is 40 at 8000~{\AA} near the \FeII\ lines.
Typical exposures were 3600 seconds long, with a total exposure time of 25,800
seconds for both the blue and red arms of the spectrograph, and a typical seeing of 0.9$"$.
The slit widths used were $0.8"$ and 1.0$"$ wide.

The raw 2D echelle spectra were reduced using the Common Pipeline
Language UVES pipeline  (version 4.2.3). After performing basic steps like bias
subtraction and flat-fielding, the pipeline optimally extracts the
quasar spectra and their 1-$\sigma$ statistical error spectra by
modeling the spatial profile of the quasar flux on the 2D frames as a
Gaussian. The same profile was used to optimally extract the ThAr
calibration frames and an optimized ThAr line-list \cite{MurphyM_07b} 
was used to determine a wavelength solution for the
quasar frames. The extracted quasar flux was corrected for the
spectrograph's blaze function using the flat-field frames, ensuring
that the flux in overlapping portions of neighbouring echelle orders
had approximately the same shape. The extracted flux from all echelle
orders, from all exposures, was placed in a vacuum--heliocentric wavelength scale and
combined with inverse-variance
weighting and a cosmic ray/bad pixel rejection algorithm using the
custom-written code {\sc uves\_popler}\footnote{Written and maintained
by MTM at \url{http://astronomy.swin.edu.au/~mmurphy/UVES_popler}.}. This
code also performed an automated continuum fit which was subsequently
improved in some sections with a manual polynomial fitting routine.
The final spectrum of HE 2243$-$60 (Figure~\ref{fig:uves})
has a wavelength coverage of
3080$-$10,430~{\AA} with two gaps between 8528$-$8663~{\AA} and
10,256$-$10,261~{\AA}. It was re-dispersed to a
velocity dispersion of 2.0~km/s/pixel and has a resolving power of $R\sim50,000$.

\subsection{Metallicity and Kinematics of the absorbing gas}

We present here our analysis of 
 the absorbing gas properties (kinematics, metallicity, dust properties) using the low-ionization ions 
and the knowledge of the systemic redshift of the host.  
The exquisite high SNR VLT/UVES spectra combined with the  galaxy properties revealed by the VLT/SINFONI data have enhanced dramatically our ability to interpret the physical properties of the gas associated with the host galaxy.

\begin{figure}[ht]

\centering
\subfigure[]{
\includegraphics[width=6cm,height=11cm]{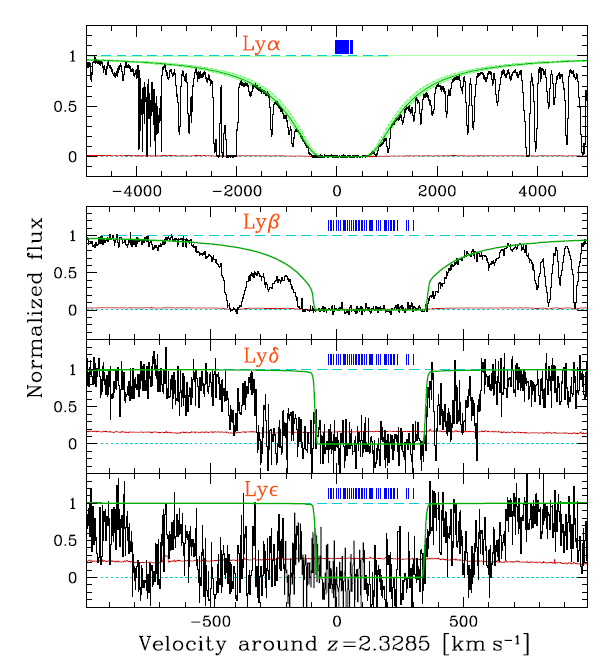}
}
\subfigure[]{
\includegraphics[width=6cm]{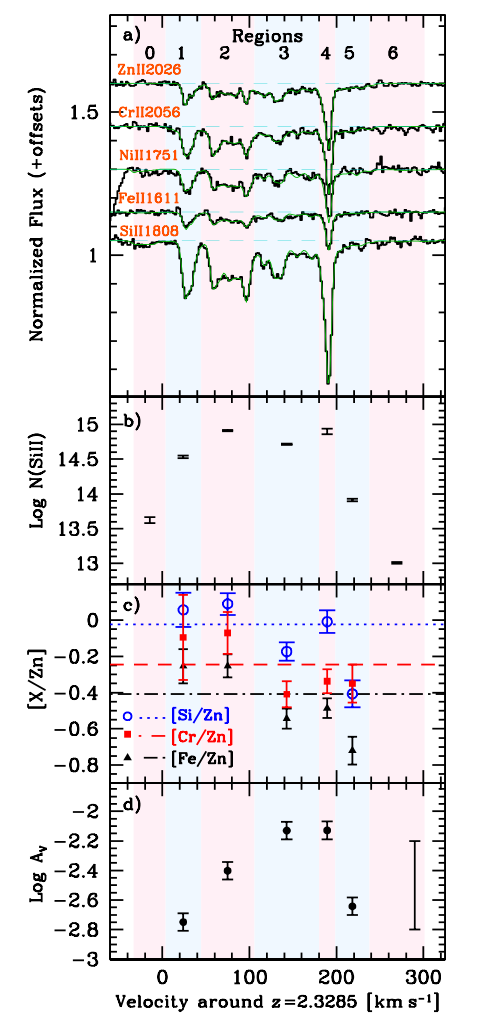}
}
\caption{{ High-resolution VLT/UVES spectrum of HE2243$-$60.}\label{fig:uves}
{\bf a} Spectrum covering the Lyman series.
We simultaneously fit the \Ly, \Lyb, \Lyd\ and \Lye\ absorption in the UVES spectrum using {\sc vpfit} by scaling the \SiII\ column densities of each of the components and keeping the same $b$-parameter values. 
Lyman series lines we detect are all heavily saturated, which prevents us
from measuring the \HI\ column density in the sub-regions (1 to 6) directly.
{\bf b} Spectrum covering the low-ionisation ions.
In panel a we show the low-ionisation ions (\ZnII, \CrII, \FeII, \SiII, \NiII) associated with the  DLA host galaxy.  
We fit each ion with multiple components using  {\sc vpfit} and identified 7 regions (labeled 0 to 6) that can be distinguished kinematically.
The galaxy systemic redshift corresponds to $0$~\kms.
Panel b  shows the \SiII\ column density profile across the seven regions.
The gas is distributed roughly equally between the components 0--3 and 4--6. 
From the ion ratios  ([Si/Zn], [Cr/Zn], [Fe/Zn]) (panel c) and 
the corresponding gas extinction $A_V$ (panel d), 
we see an increasing amount dust depletion
towards the components 3 and 4.  Error bars show the measurement errors
and the systematic uncertainty is shown at $v=$290~\kms.}
\end{figure}

The VLT/UVES spectrum of HE 2243$-$60  is shown in Figure~\ref{fig:uves}a for the Lyman series and
in Figure~\ref{fig:uves}b for some of the low-ionization ions.
We fit all the metal lines (\ZnII, \CrII, \FeII, \SiII, \NiII) with multiple components
using  {\sc vpfit} version 9.5\footnote{{\sc vpfit} is maintained by R.F.~Carswell at \url{http://www.ast.cam.ac.uk/~rfc/vpfit.html}.}.  The total \SiII\ and \ZnII\ column densities are $\log N(\SiII)[\cmsq]=15.41\pm0.05$ and $\log N(\ZnII)[\cmsq]=12.53\pm0.05$ and the other elements are listed in Table~\ref{table:uves}.
%{\it The blue tick marks show the central position of each component and the green line represents the fit to the spectra.}
We identified 7 regions (labeled 0 to 6) that can be distinguished kinematically.
For instance, there is a strong component at $\sim180$~\kms\ and a number of sub-components with velicities ranging
from 20 to 160~\kms. Figure~\ref{fig:uves}b (panels a, b) shows that the \SiII\ gas column density is contained mostly in components 2 and 4, where each have  approximately 50\%\ of the total column density.

We simultaneously fit the \Ly, \Lyb, \Lyd\ and \Lye\ absorption in the UVES spectrum (Figure~\ref{fig:uves}a) using {\sc vpfit} by scaling the \SiII\ column densities of each of the {\sc vpfit} components and keeping the same $b$-parameter values.  We find the total \HI\ column density to be 
%\begin{eqnarray}
$\log \NHI[\cmsq]=20.62\pm0.05$, well above the 20.3 threshold value for DLAs, indicating that the gas is likely to be almost entirely neutral.
%\end{eqnarray}
This column density is consistent with the value, 20.67, from the earlier spectra
\cite{LopezS_02a}. In our measurement, the main source of error is from the continuum uncertainty around the \Ly\ through. Unfortunately, the VLT/UVES spectrum does not allow us to estimate the \HI\ column in each of the components, preventing us from performing a photo-ionization analysis, because the Lyman series lines we detect are all heavily saturated.  However, under the conservative assumption of constant metallicity across the profile, we use the \SiII\ column density profile as a proxy for the \HI\ distribution among the fitted absorption components, and estimate that $\sim$50\%\ of the total \HI\ column density of $\log\NH\simeq 20.6$ is contributed by components 1--3.

Using the \HI\ column and the refractory element least depleted on dust grains \Zn, the global metallicity is estimated to be 
\begin{equation}
[\ZnH]=\Zndla\pm0.05.\label{eq:Zdla}
\end{equation}
Similarly, the  \Fe\ abundance is [\Fe/H]=-1.13$\pm0.06$.  These measurements are  0.3~dex larger than
the original values \cite{LopezS_02a}  ([\ZnH]$\simeq-1.10$) based on a shallower spectrum. 
For DLAs, this metallicity range is in the upper quartile of samples of DLAs \cite{KulkarniV_02a,ProchaskaJ_03d,KhareP_07a,EllisonS_12a,RafelskiM_12a}.
More importantly, for the purpose of understanding this galaxy-quasar pair, 
we find that 
 the gas metallicity at 26~kpc is much smaller than that of the ISM of the host,  which is $\sim 1/3$ $Z_{\odot}$
(Eq.~\ref{eq:Z}).
Taking this comparison at face-value, this decrease of $\sim$0.5~dex in metallicity over 26~kpc
 excludes the possibility that the gas probed by the QSO line-of-sight traces entrained material in an outflow.
Indeed, in such a scenario the outflow metallicity would be either enriched with metals \cite{MartinC_02a,TrippT_11a} or be that of the ISM.

We now turn to our analysis of the low-ionization ions across the profile.
Figure~\ref{fig:uves}b (panel c) shows the line ratios [\Fe/\Zn], [\Cr/\Zn], and [\Si/\Zn] as
triangles, squares and circles, respectively. Overall, one sees little depletion for  \Si\ compared to \Zn,
and that \Cr\ and \Fe\ are more depleted, as one might have expected.  
More interestingly, the  three ion ratios show a clear trend with velocity where
the regions at the highest velocities are more depleted. From these line ratios,
we use the correlation between extinction $A_V$
and dust-phase metallicity \cite{VladiloG_06a}
to estimate the total extinction $A_V$.
The resulting $A_V$ profiles (Figure~\ref{fig:uves}b, panel d) shows that  most of the dust is located in the components with velocities $v=$150--200\kms. Similarly,
using the Jenkins method \cite{JenkinsE_09a}, one can extract a constraint on $\NHI+\log Z$, the sum of the metallicity and the gas column.
We find that this quantity follows the derived $A_V$ values very closely.

This dust profile shown in Figure~\ref{fig:uves}b would mean that the gas metallicity is higher for component no. 4 than
for the gas with small line-of-sight velocities (components no 1 to 3),
assuming a constant dust-to-metal ratio. Thus, if we
assume that the metallicity is lower in the low velocity components (no 1 to 3), 
this would imply that the
contribution to the total \NH\ from the accreting components will be larger than the 50\%\ value because the
$N(\Si)$ is similar for the two groups and that the estimated accretion rate should be revised
upwards.

Finally, we note that the high-ionization ions (e.g. \CIV, \OVI) 
associated with this system have a kinematic profile \cite{FoxA_07a,FoxA_07b} that is very different: contrary to the low-ionization gas, it is symmetric around the galaxy systemic velocity.  These could be the signatures of the virial shocks.

\clearpage

\section{Kinematic modelling}
\label{section:model}

\begin{figure}[ht]
\centering
\includegraphics[width=16cm]{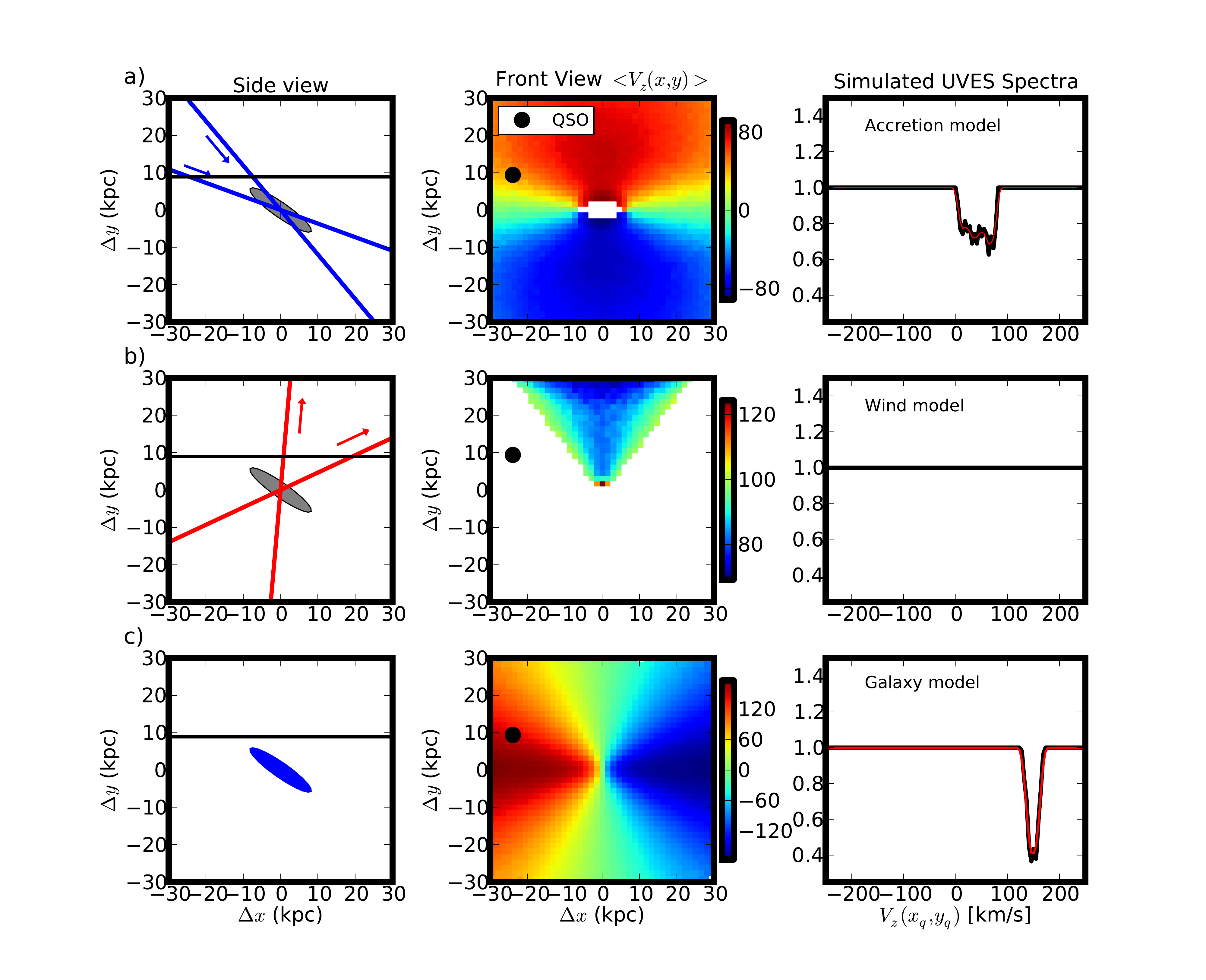}

\caption{{Simulated line-of-sight kinematics for various components.}\label{fig:model}
The accretion, outflow and disk components  are shown in {\bf a}, {\bf b} and {\bf c},
respectively.
For each components of the modeling (disk, outflow, inflow), 
we show a side view (left), a front view of the average line-of-sight velocity (middle)
and a simulated spectra at the UVES resolution (right).
For typical opening angles of $60$~$\degree$, a bi-conical outflow does not intercept the sight-line.
A simple accretion model with a non-zero radial component can account for the intermediate velocities seen
in the VLT/UVES spectra (components 1-3 in Figure~1, \ref{fig:uves}) between 0 and 100 \kms.
}

\end{figure}

Given our tight geometric constraints on the quasar apparent location with respect to the galaxy, we can simulate UVES profiles for various physical scenarios using simplified toy models.
We use a Monte Carlo approach where typically $10^5$ `clouds' are created to fill these geometric configurations with simplified kinematics, which allows us to generate any sight-line absorption profile  
for any quasar location.   Because of the random nature of the Monte Carlo approach, there will be irregularities in the simulated profiles (Figures 1d, ~\ref{fig:model}).  The absorption profile is generated with velocity bins given by the UVES pixel size  and convolved with the instrument Line-Spread-Function.

Figure~\ref{fig:model} shows three such toy models, namely that of 
accretion, a bi-conical wind, and disk, in a, b, and c, respectively. The left (middle) panels in Figure~\ref{fig:model} show a side (front) view of the toy models, respectively.
The middle panels show the average line-of-sight $V_z$ velocity and the right panels show
the simulated UVES absorption profiles for our quasar-galaxy configuration. 

For the disk model, we assume a flat rotation curve,
for the outflow model, we assume a constant radial outflow speed, and for the accretion model
a constant inflow speed (radial) representing a non-zero mass flux.
In all cases, we assume an axis symmetry around the rotation axis.
For the accretion and outflow, we ensure mass conservation.
While the true physical situation is likely to be more complex, this exercise shows that one can
reproduce the kinematics of components no 1 to 3 with an accreting gas, while the outflow model
fails to account for the kinematics of the low velocity gas unless the cone opening angle is $>140$~$\degree$, which is not consistent with current constraints at $z=0.01$--1 on the cone opening angle \cite{BordoloiR_11a,BoucheN_12a,KacprzakG_12b}.  Moreover, an outflow occurring in such a wide cone will lead to a kinematic profile very different from the one observed: it will have a peak at some velocity (say 200~\kms) given by the outflow speed  and a continuous decrease to about 100~\kms, leaving the components at 0 to 100~\kms, unaccounted for.

For the accretion model, one may use various geometric configurations, such a flat cylindrical, a torus-like cone (Figure~\ref{fig:model}a), or a disk-like geometry (Figure~1d) as these lead to similar kinematic profile.  
One may   want to relax our assumption of radial orbits  as the material
is more likely to spiral inwards. This will introduce  a small correction in Eq.~1, a correction of order unity $\mathcal{O}(1)$ from the sine or cosine projection factor from the spiral opening angle. Hence, the resulting accretion rate could be smaller by a factor of two to three.
Nonetheless, our accretion toy model represents halo gas that could be spread over 10--15~kpc (above and below the disk).  This situation is similar to local examples.
Indeed, in local galaxies, such as NGC 891, 21~cm observations \cite{OosterlooT_07a} have revealed large amounts of extra-planar neutral gas in the halos of galaxies, with scale heights as large as 20~kpc. In these local examples, an inflow component of low-angular momentum gas is usually required in order to account for the kinematics seen in the 21~cm IFU data \cite{FraternaliF_08a}.

\pagebreak
%%%%%%%%%%%%%%%%%%%%%FIGURES

%\begin{figure}
%\centering
%\includegraphics[width=9.5cm]{Figure_S1.png}
%\caption{{ The $K$-band SINFONI field.} The greyscale represents a narrow band image (rest-frame \Ha) with
% the continuum subtracted.
%The QSO HE2243$-$60 (cross) and the host galaxy ('G') are marked. 
%The residuals from the continuum subtraction are visible both near the QSO and near
%the  position labeled 'R'.  This secondary location for the QSO (with negatif fluxes) is %caused by the
%sky-subtraction technique used in the near-IR.  
%\label{fig:sinfo}}
%\end{figure}

%\pagebreak

\pagebreak

\pagebreak

\pagebreak

\pagebreak

\pagebreak

\clearpage

%%%%%%%%%%%%%%%%% TABLES

\begin{table*}
\centering
\caption{Journal of the VLT/SINFONI Observations. \label{table:observations}}
\begin{tabular}{llllllll}
Instrument & Mode		 	&	Grating	& Pixel 	& $t_{\rm exp}$	&	PSF FWHM	 &   Run ID		\\
 & & & (\arcsec) & (s) & (\arcsec) & \\
\hline

SINFONI &NGS	 	&	K	& 0.125$\times$0.25	&   $12000$		&  0.25	& 383.A-0750 \\
SINFONI &NGS		&	H	& 0.125$\times$0.25	&   $12000$		&  0.4 &	 383.A-0750 \\
SINFONI &NGS	        &       J        & 0.125$\times$0.25 &       $ 10800$ & 0.65  & 088.B-0715\\
\hline
%UVES & & & & & & & 065.A-0411 \\
%UVES & & & & & & & 074.A-0201 \\
%UVES & & & & & & & 067.A-0567 \\
%\hline 
\end{tabular}
\end{table*}

\clearpage

\begin{table}
\centering
\caption{{ Summary of the host emission properties}.
 \label{table:param}\label{table:flux}}
\begin{tabular}{llll}
\hline 
Line &    Flux (\flux)  \\
 &   ($\times 10^{-17}$) &    \\
\hline 
\Ha~$\lambda$6564   & 9.1(0.2)   \\
\NII~$\lambda$6583     & $<$1.22 \\
\Hb~$\lambda$4963     & 1.7(0.4)\\
\OI~$\lambda$6300  & $<$1.15 \\
\OII~$\lambda$3727     & 3.8(0.4) \\
\OIII~$\lambda$5007    & 8.1(0.2) \\
\hline
\end{tabular}
\end{table}

\clearpage

\begin{table*}
\small
\centering
\caption{{ Summary of the host kinematics properties}.
We use two or three-dimensional fits on the \Ha\ data cube to determine the galaxy size and kinematic
parameters. We use an exponential flux profile and for the 3D fits
the rotation curve $f_V(r)$ is assumed to be either a rising exponential (`Exp') or an arctangent ('arctan').
 \label{table:param}\label{table:kinematics}}
\begin{tabular}{llllllllllll}
\hline 
  Fit & $f_V(r)$   & $R_d$   & $i$   & PA  &  $r_p$ & $V_{\rm max}$ & $\sigma_i$   \\
     & &(kpc)   & (deg) & (deg) &    & (\kms) & (\kms)   \\
\hline 
 2D& n.a. &   2.30(0.15) & 54(4) &  62(6) & $\cdots$ & $\cdots$& $\cdots$  \\
 3D&arctan & 2.37(0.05) & 55(1) & 54(1) & 1.22(0.5) & 145(16) & 157(5)      \\
 3D&Exp. & 2.30(0.05) &  53(1) & 55(1) & 3.5(0.6) & 180(32) & 158(5)    \\
\hline
\end{tabular}
\end{table*}

\clearpage

\begin{table}
\centering
\caption{{ Column densities and Abundances Measurements}.}
\begin{tabular}{llll }
X & $\log N_{\rm X}$ & [X/H] & Reference	  \\
(1)    &  (2)  &  (3)   & (4)  \\
\hline
H  &  20.62 $\pm$0.05 &  	N.A. & This work   \\
CrII &  13.31 $\pm$0.01 & -0.96$\pm$0.06 & This work\\
FeII &  14.96 $\pm$0.01 & -1.13$\pm$0.06 & This work \\
NiII & 13.88 $\pm$0.01 & -0.96$\pm$0.05 & This work\\
SiII &  15.41 $\pm$0.01 & -0.74$\pm$0.07 & This work\\
ZnII &  12.53 $\pm$0.01 & \Zndla$\pm$0.05 & This work\\
 O  & 16.21 &-0.92$\pm$ 0.22  &  \cite{LopezS_02a} \\
OVI & 14.98 &  $\cdots$  & \cite{FoxA_07a}  \\
CIV& 14.73  &  $\cdots$ & \cite{FoxA_07a} \\
NV & 13.50 &  $\cdots$ & \cite{FoxA_09a} \\
NI & 14.59 &  $\cdots$ & \cite{FoxA_09a} \\
%\hline
%$\log Z$  &  & -1.84$\pm$0.08  &  & &This paper\\
%$F_{\star}$ &   & -1.15$\pm$0.07 &  & &This paper\\
\hline
\end{tabular}\\
(1) Element name;
(2) Column density;
(3) Abundances;
(4) References.
\label{table:uves}
\end{table}

%\bibliographysec{references}

%%%%%%%%%%%%%%%%%%%%%%%%%%%%%%%%%%%%%%%%%%%%%%%%%%%%%%%%%%%%%%%%%%

\end{document}